\documentclass[preprint,12pt,3p]{elsarticle}
\usepackage{lineno,hyperref}
\usepackage{amsmath}
\usepackage{graphicx}
\usepackage{epstopdf}
\usepackage{float}
\usepackage{multirow}
\usepackage{hhline}
\usepackage{xcolor}
\usepackage{subcaption}
\usepackage{lineno}
\usepackage[utf8]{inputenc}
\usepackage{soul}
\usepackage{setspace}
\usepackage[normalem]{ulem}

\makeatletter
\newcommand{\vast}{\bBigg@{4}}
\newcommand{\Vast}{\bBigg@{5}}
\def\ps@pprintTitle{%
	\let\@oddhead\@empty
	\let\@evenhead\@empty
	\let\@oddfoot\@empty
	\let\@evenfoot\@oddfoot
}
\makeatother
\begin{document}
\begin{frontmatter}

\title{Scintillator light yield measurements with waveform digitizers}
\author[1]{T.A. Laplace\corref{cor1}}
\ead{lapthi@berkeley.edu}
\cortext[cor1]{Corresponding author}
\author[1,2]{B.L. Goldblum}
\author[1]{J.A. Brown}
\author[1]{J.J. Manfredi}

\address[1]{Department of Nuclear Engineering, University of California, Berkeley, California 94720 USA}
\address[2]{Nuclear Science Division, Lawrence Berkeley National Laboratory, Berkeley, California 94720 USA}

\begin{abstract}
The proton light yield of organic scintillators has been measured extensively in recent years using fast waveform digitizers and large discrepancies exist in the values reported by different authors. In this letter, we address principles of digital signal processing that must be considered when conducting scintillator light yield measurements. Digitized waveform pulse height values are only proportional to the amount of scintillation light if the temporal shape of the scintillation pulse is independent of the amount of energy deposited. This is not the case for scintillation pulses resulting from fast neutron interactions in organic scintillators. Authors measuring proton light yield should therefore report pulse integral values and ensure that the integration length is long enough to capture most of the scintillation light. 
\end{abstract}


\end{frontmatter}

The absolute light yield of a scintillator can be defined as the number of photons generated per unit energy deposited by a particle slowing down in the scintillating medium \cite{Holl}. The term \textit{light yield} is often referred to in the literature as a ratio of the  number of scintillation photons to the energy deposited \cite{SYSOEVA2002}, but has also been used interchangeably with the terms \textit{light output} and \textit{scintillation response} to represent a quantity proportional to the number of scintillation photons \cite{Laplace, birks1964-55, BROOKSDevelopment}. 

For fast neutron detection using organic scintillators, the primary mechanism for light generation is n-p elastic scattering, and it is the recoiling proton that deposits energy resulting in molecular excitation and ionization. For the purposes of this discussion, the \textit{proton light yield} is defined as the number of scintillation photons generated by a recoiling proton in an organic scintillator. By extension, the proton light yield relation is defined as the relationship between the amount of scintillation light generated by a recoiling proton as a function of its energy deposited in the medium. This relationship varies non-linearly with recoil energy due to quenching phenomena \cite{birks1964-187}. The potential use of organic scintillators as neutron detectors in various arms control and nuclear security applications has prompted efforts to precisely characterize the proton light yield of a variety of organic scintillators in the energy range relevant for fission neutrons. This renewed interest occurred coincidently with the broad availability of fast waveform digitizers. 
 
Representative traditional analog chains for measurement of the proton light yield of organic scintillators coupled to photomultiplier tubes (PMTs) are detailed in Refs.~\cite{Verbinski, Smith1968, Iwanowska-309Analog}. Two signals are output from the PMT: the signal from the anode is fed to an analog timing chain and the signal from one of the dynodes is fed to a preamplifier followed by a shaping amplifier. The shaping time plays a crucial role in estimating the collected charge. When the shaping time is much less than the charge collection time, the voltage signal output of the shaping amplifier reproduces the current input. This case is recommended for timing information or when high event rates are prioritized over accurate energy information~\cite{Knollthis}. If the shaping time is much greater than the charge collection time, the shaping amplifier will act as a current integrator. The maximum voltage output (i.e., amplitude of the signal) will be proportional to the charge generated within the PMT as long as the load circuit capacitance remains constant~\cite{Knollthis}. For scintillator signals, the maximum voltage of the output of the shaping amplifier is referred to as the ``pulse height.'' With a sufficiently long shaping time, this corresponds to the integral of the PMT charge, which is proportional to the number of scintillation photons (assuming a linear PMT response \cite{friendLinearity}). For example, in their pioneering measurement of the proton light yield of the NE-213 organic liquid scintillator, Verbinski et al.\ used a 2.4~$\mu$s shaping time \cite{Verbinski}. This is long in comparison to typical temporal responses of organic scintillator pulses on the order of $1-100$~ns \cite{BollingerThomas,BrownBibenzyl}. 
 
Since the early 2000s, the availability of fast waveform digitizers has enabled a different approach to signal acquisition in nuclear instrumentation. Instead of feeding the PMT output to analog modules to accomplish various operations, the signal is directly fed to the digitizer, where full waveforms are digitized and then stored for offline signal processing. This allows for fine-tuning of various signal processing operations such as integration length, pulse-shape discrimination parameters, baseline estimation, pile-up rejection, etc. Measurements of the proton light yield of organic scintillators using waveform digitizers have yielded large discrepancies between different groups. Brown et al.~\cite{brown2018proton} demonstrated that differences in proton light yield measurements of EJ-309 resulted in part from the chosen integration length of the digitized waveforms. The proton light yield has been reported in several publications by estimating the maximum of the scintillation pulse (i.e., the digitized waveform pulse height) as opposed to the pulse height from the output of the shaping amplifier in an analog signal chain~\cite{enqvist,lawrence,wang,ShinOGlass2019}. Such analyses reporting proton light yield obtained using digitized waveform pulse heights carry several flaws, detailed below.

First, the temporal shapes of pulses resulting from electron and proton recoils in organic scintillators are different. For some materials, the pulse shape is sufficiently different such that a simple method like comparing the amount of light from the tail of the pulse to the total scintillation light allows for separation of events generated by different types of recoiling particles, i.e., pulse shape discrimination (PSD) via a charge integration approach~\cite{owen1958PSD,brooks1956organic}. The proton light yield is often reported in units of MeV electron-equivalent (MeVee), which corresponds to the equivalent amount of scintillation light produced by recoiling electrons of a given energy. For a given number of scintillation photons, the average pulse originating from a neutron interaction has less prompt light (due to increased ionization quenching) and more delayed light (due to increased triplet-triplet annihilation) than a pulse resulting from a $\gamma$-ray interaction. The use of the digitized waveform pulse height to obtain the proton light yield in MeVee therefore results in an underestimation of the proton light yield relation.

\begin{figure}
	\center
	\includegraphics[width=0.80\textwidth]{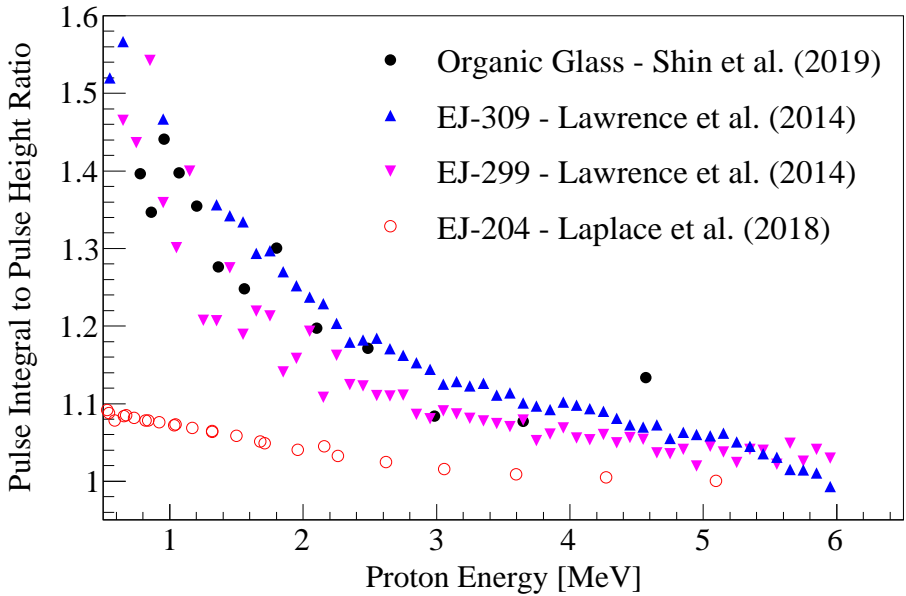}
	\caption{Ratio of the proton light yield obtained using the pulse integral relative to the digitized waveform pulse height as function of proton recoil energy. Note the zero-suppressed axes. \label{PIDVSPHD}}
\end{figure}

Second, the proton pulse shape is energy dependent. As a result, digitized waveform pulse heights are not proportional to the number of scintillation photons as function of energy. Figure~\ref{PIDVSPHD} shows the ratio of the proton light yield obtained using the integral of the digitized waveform to the height of that waveform for the same datasets for four different organic scintillators. Shin et al.\ examined an organic glass scintillator using an integration length of 150~ns to ensure $>95\%$ of the scintillation light was captured, and therefore the pulse integral value is expected to be proportional to the amount of scintillation light generated in the detector~\cite{ShinOGlass2019}. Lawrence et al.\ used a 180~ns integration window for the liquid EJ-309 and  plastic EJ-299-33 organic scintillators~\cite{lawrence}. Although measurements of the temporal response of the scintillation light generated by EJ-309 are not available in the literature, we estimated this integration length as capturing $\sim$90\% of the scintillation light by examining average digitized waveforms in the MeV proton energy range. In the case of non-PSD scintillators, data adapted from previous measurements of the EJ-204 proton light yield are provided, which employed an integration length of 200~ns to capture at least 95\% of the scintillation light within the acquisition window~\cite{Laplace}. The ratio shown in Fig.~\ref{PIDVSPHD} varies as a function of proton recoil energy for all materials examined, demonstrating that the two quantities are not proportional. For EJ-204, the ratio remains energy dependent, but to a lesser degree than that of the PSD materials. This variation suggests a change in the temporal response of these organic scintillators as function of proton energy.

\begin{figure}
	\center
	\includegraphics[width=0.80\textwidth]{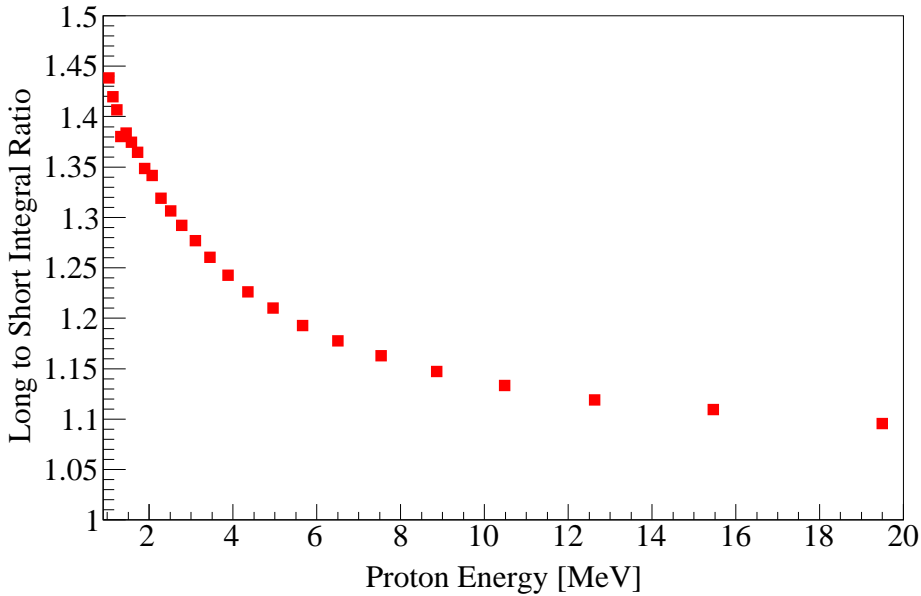}
	\caption{Ratio of the proton light yield of EJ-309 from Brown et al.\ \cite{brown2018proton} using a long (300~ns) to short (30~ns) integral as a function of proton recoil energy. Note the zero-suppressed axes. \label{LIVsSI}}
\end{figure}

Figure~\ref{LIVsSI} shows a comparison of the EJ-309 proton light yield obtained using a long (300~ns) and short (30~ns) integration length from Brown et al.~\cite{brown2018proton}. Similarly, the ratio is energy dependent, suggesting that a short integration length is not proportional to the total scintillation light. While the prompt fluorescent light is captured by a 30~ns integration length, only a fraction of the delayed light is integrated. The ratio of prompt-to-delayed light is dependent upon the nature and energy of the incident particle, the degree of ionization quenching in the medium, and the resulting density of excited and ionized molecules.

Pulse height data may be used in an attempt to recover information in high count rate experiments where scintillator pulses overlap. Algorithms have been developed and demonstrated for slow inorganic scintillators to deconvolve digitized waveforms with pulse pileup~\cite{guo2005}, though in this case knowledge of the temporal response of the scintillation light was required. The complexity is increased with organic scintillators, given a pulse shape that is both particle and energy dependent. 

In summary, the shape of scintillation pulses resulting from neutron interactions in organic scintillators is dependent upon the energy deposited. Therefore, the proton light yield cannot be measured using the maximum pulse height of a digitized waveform. This value will be proportional to the peak photon fluence on the photocathode and not to the number of scintillation photons created. Authors measuring proton light yield should report pulse integral values and ensure that the integration length is long enough to capture most of the scintillation light. 

\section*{Acknowledgements}
This work was performed under the auspices of the U.S. Department of Energy by Lawrence Berkeley National Laboratory under Contract DE-AC02-05CH11231 and is based upon work supported by the U.S. Department of Energy, National Nuclear Security Administration, Office of Defense Nuclear Nonproliferation Research and Development (DNN R\&D) via the Nuclear Science and Security Consortium under Award Number DE-NA0003180.

\section*{References}

\bibliography{./PHVsPI}
\bibliographystyle{unsrtnat}

\end{document}